\documentclass[aps,pra,twocolumn,superscriptaddress]{revtex4}

\usepackage{amsfonts,epsfig}
\usepackage{latexsym}
\usepackage{amsmath}
\usepackage{amssymb}
\usepackage{amsfonts}
\usepackage{amsthm}
\usepackage{mathrsfs}
\usepackage{natbib}
\usepackage{color,verbatim,graphics}
\usepackage{psfrag}
\DeclareMathAlphabet{\mathrsfs}{U}{rsfs}{m}{n}
\DeclareMathAlphabet{\mathpzc}{OT1}{pzc}{m}{it}
\DeclareMathAlphabet{\matheus}{U}{eus}{m}{n}
\DeclareMathAlphabet{\mathbbold}{U}{bbold}{m}{n}

\newcommand{\HH}{\mathcal{H}}
\newcommand{\CC}{\mathbb{C}}

\newcommand{\ba}{\begin{eqnarray}}
\newcommand{\ea}{\end{eqnarray}}
\newcommand{\ban}{\begin{eqnarray*}}
\newcommand{\ean}{\end{eqnarray*}}
\newcommand{\Tr}{\operatorname{Tr}}

\newcommand{\ket}[1]{|#1\rangle}
\newcommand{\bra}[1]{\langle#1|}
\newcommand{\braket}[2]{\langle#1|#2\rangle}

\newcommand{\etal}{{\it{et al.~}}}

\begin{document}

\title{Device-independent certification of entangled measurements}

\author{Rafael Rabelo}
\affiliation{Centre for Quantum Technologies, National University of Singapore, 3 Science drive 2, Singapore 117543}
\author{Melvyn Ho}
\affiliation{Centre for Quantum Technologies, National University of Singapore, 3 Science drive 2, Singapore 117543}
\author{Daniel Cavalcanti}
\affiliation{Centre for Quantum Technologies, National University of Singapore, 3 Science drive 2, Singapore 117543}%
\author{Nicolas Brunner}
\affiliation{H.H. Wills Physics Laboratory, University of Bristol, Tyndall Avenue, Bristol, BS8 1TL, United Kingdom}
\author{Valerio Scarani}
\affiliation{Centre for Quantum Technologies, National University of Singapore, 3 Science drive 2, Singapore 117543}
\affiliation{Department of Physics, National University of Singapore, 2 Science Drive 3, Singapore 117542}



\begin{abstract}
We present a device-independent protocol to test if a given black-box measurement device is entangled, that is, has entangled eigenstates. Our scheme involves three parties and is inspired by entanglement swapping; the test uses the Clauser-Horne-Shimony-Holt (CHSH) Bell inequality, checked between each pair of parties. Also, focusing on the case where all particles are qubits, we characterize quantitatively the deviation of the measurement device from a perfect Bell state measurement.
\end{abstract}


\maketitle



\textit{Introduction} - The concept of device-independent information processing relies on performing reliable information tasks on untrustworthy apparatuses. This idea first showed its importance in quantum key distribution protocols for cryptography, where security and privacy could be tested and assured even in the hypothetical case that the cryptographic devices are provided by some malevolent third party \cite{ekert,diqkd} (see also \cite{diqkd2}). Other tasks were also generalized to the device independent scenario, such as random number generation \cite{colbeck,dirandom} and quantum state estimation \cite{bardyn}, the latter starting an alternative approach to self-testing as initially proposed by Mayers and Yao \cite{my}.

Inspired by this line of research, Navascu\'es and V\'ertesi \cite{miguel-tamas} have recently introduced the task of testing if a measurement device that acts on a bipartite quantum system is \textit{entangled}, that is, if at least one of its eigenstates is not separable (or, more generally, its POVM elements do not factor in the subsystems). However, their solution cannot be considered as device-independent, since further assumptions are required. In this letter, we present a device-independent realization of this task.

Specifically, in order to show that a measurement is entangled, we are going to assess that it is \textit{entangling} in an entanglement swapping scenario \cite{entgswap}. Suppose $A$ and $B$ are initially not entangled; rather, $A$ is initially entangled with a system $C_A$ and $B$ with a system $C_B$. Then, if a measurement on $C_A-C_B$ creates entanglement between $A$ and $B$, that measurement was entangled. To perform a device-independent test means that we cannot \textit{assume} this scenario, but rather, we want to \textit{certify a posteriori} that swapping has indeed happened. Entanglement is checked in a device-independent way using Bell's inequalities. Here we shall use only the Clauser-Horne-Shimony-Holt (CHSH) inequality
\ba
S&=&E_{11}+E_{12}+E_{21}-E_{22}
\ea
where $E_{xy}=\textrm{prob}(a=b|x,y)-\textrm{prob}(a\neq b|x,y)$, and $\textrm{prob}(a,b|x,y)$ is the joint probability of obtaining results $a$ and $b$ given that measurements $x$ and $y$ were performed.

It is important to stress that the only assumption that may go into our test is the existence of two clearly defined subsystems in Charlie's hands (we shall see later that it can be considered very natural in some implementations, and that it can be entirely dispensed with in the idealized cases that we study below). No other assumption needs to be made: for instance, the state could be of arbitrary Hilbert space dimension and could be entangled along any partition.





\textit{Protocol} - For each run, Alice chooses at random one out of two measurements $A_1$ or $A_2$ with binary outcome $a_i\in\{-1,+1\}$; Bob, one of two measurements $B_1$ or $B_2$, also with binary outcome $b_j\in\{-1,+1\}$; Charlie, one out of three measurements $C_1$, $C_2$ or $C_3$, with four outcomes $c_k\in\{1,2,3,4\}$. The goal is to \textit{guarantee that $C_3$ is an entangled measurement}.

On the statistics resulting from a large number of repetitions, the following tests are performed:
\begin{enumerate}
\item The cases where Charlie has measured $C_1$ or $C_2$ are used to test the CHSH inequality both with Alice ($S_{AC}$) and with Bob ($S_{BC}$). For this, Charlie has to define a classical processing that transforms his four outcomes into two bits, one to be correlated with Alice and one with Bob.
\item When Charlie has measured $C_3$, Alice and Bob check CHSH among themselves, obtaining four numbers $S_{AB|c_3}$ conditioned on the result $c_3$ obtained by Charlie:
\ba
\begin{array}{lcl}
S_{AB|1}=-S_{AB|4}&=&E_{11}+E_{12}+E_{21}-E_{22}\,,\\
S_{AB|2}=-S_{AB|3}&=&E_{11}+E_{12}-E_{21}+E_{22}\,.
\end{array}\label{versions}
\ea
Note that Alice and Bob do not need to know $c_3$ in each run, since their measurement settings are always the same. The statistics (\ref{versions}) can be checked at the end of the whole experiment.
\end{enumerate}

Now we are going to show that this protocol can lead to a device-independent test of the fact that $C_3$ is entangled. On the one hand, notice that in quantum physics it is possible to achieve \ba
S_{AC}=S_{BC}=S_{AB|c_3}=2\sqrt{2}&& \textrm{for all $c_3\in\{1,2,3,4\}$}\nonumber\\
\label{cond2}
\ea
already in a four qubit scenario. The system is prepared in the state $\ket{\Phi^+}_{AC_A}\otimes \ket{\Phi^+}_{BC_B}$. Alice's and Bob's measurements are: $A_1=Z$, $A_2=X$, $B_1=(Z+X)/\sqrt{2}$, $B_2=(Z-X)/\sqrt{2}$. Charlie's measurements are $C_1=(Z+X)/\sqrt{2}\otimes Z$, $C_2=(Z-X)/\sqrt{2}\otimes X$ and $C_3$ is the Bell-state measurement, in which outcome $c_3\in\{1,2,3,4\}$ indicates the projection on one of the Bell states $\ket{\Phi_1}=\ket{\Phi^+}$, $\ket{\Phi_2}=\ket{\Phi^-}$, $\ket{\Phi_3}=\ket{\Psi^+}$, $\ket{\Phi_4}=\ket{\Psi^-}$.

On the other hand, the observation that either $S_{AC}$ or $S_{BC}$ is \textit{exactly} $2\sqrt{2}$ guarantees in a device-independent way that (i) $\rho_{AB}$ is separable and (ii) Charlie's system can be seen as composite of two subsystems. Indeed, if (say) $S_{AC}=2\sqrt{2}$, up to local isometries the three-partite state is $\Phi_{AC}\otimes\rho_{A'BC'}$, where $\Phi=\ket{\Phi^+}\bra{\Phi^+}$, while $A'$ and $C'$ represent additional degrees of freedom of Alice and Charlie that may even be entangled among themselves and with Bob's system, but are not involved in the measurements \cite{bmr92,pr92,mckague}. So, Charlie has two uncorrelated subsystems, $C$ and $C'$. Moreover, if $S_{AB|c_3}>2$ is observed for any value of $c_3$, then $\rho_{AB|c_3}$ must be entangled, which is possible only if $C_3$ is an entangling measurement given that $\rho_{AB}$ is separable.

In summary, we have shown that
\ba
\big(S_{AC}=2\sqrt{2} \textrm{\bf{ or }}  S_{BC}=2\sqrt{2}\big) \textrm{\bf{ and }} S_{AB|c_3}>2\nonumber\\\Longrightarrow \;\textrm{$C_3$ is entangling and entangled}.\label{crit1}
\ea This is a device-independent result, since only the monogamy induced by the maximal violation of CHSH plays a role, without any \textit{a priori} assumption on the state, the Hilbert space dimension or the measurements.

The criterion (\ref{crit1}) can in principle be sharpened by exploiting both $S_{AC}$ and $S_{BC}$. The idea is that the value of $S$ provides information on the commutator between the two local measurements of each party; so, $S_{AC}$ and $S_{BC}$ constrain $[A_0,A_1]$ and $[B_0,B_1]$ respectively. In turn, something can be inferred on the Bell operator ${\cal B}_{AB}$ which, of course, determines the maximal violation achievable with separable states. Indeed, we can prove that
\ba
\big(S_{AC}=2\sqrt{2} \textrm{\bf{ and }}  S_{BC}=2\sqrt{2}\big) \textrm{\bf{ and }} S_{AB|c_3}>\sqrt{2}\nonumber\\\Longrightarrow \;\textrm{$C_3$ is entangled}.\label{crit2}
\ea Here is a sketch of the proof (see Appendix for details): one first shows that the first two conditions force the four Bell operators ${\cal B}_{AB|c_3}$ to have $2\sqrt{2}$ as eigenvalue; then, given such an operator, one shows that separable states can only reach the value $\sqrt{2}$, thus generalizing to the device-independent scenario an observation made in earlier works assuming qubits \cite{seevinck, roy}.

We have been able to derive only quantitative criteria that rely on at least one between $S_{AC}$ and $S_{BC}$ being exactly $2\sqrt{2}$. The task of relaxing this constraint is left for future work; the main difficulty arising from the fact that, even for the smallest deviation from the ideal values, $\rho_{AB}$ can't be guaranteed to be separable anymore \cite{foot1}. Similarly, one cannot guarantee anymore, in a device-independent way, that Charlie has two subsystems: as we mentioned above, this is an assumption that must be made. This assumption may however be very natural in some implementations, in which Charlie receives one quantum signal from Alice and one from Bob.


\begin{figure}
	\centering
		\psfrag{A}[][][1]{$A$}
		\psfrag{B}[][][1]{$B$}
		\psfrag{C}[][][1]{$C$}
		\psfrag{a}[][][1]{$a$}
		\psfrag{b}[][][1]{$b$}
		\psfrag{c}[][][1]{$c$}
		\psfrag{x}[][][1]{$x$}
		\psfrag{y}[][][1]{$y$}
		\psfrag{z}[][][1]{$z$}
		\psfrag{d}[][][1]{$a)$}
		\psfrag{e}[][][1]{$b)$}
		\includegraphics[width = 65mm, height = 32mm]{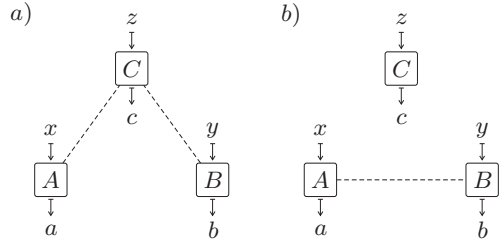}
	\caption{The scenario consists of three parties, $A$, $B$ and $C$, space-like separated, each holding a black-box device that performs measurements on quantum systems. The protocol is divided into two parts: $a)$ parties $AC$ and $BC$ evaluate the CHSH inequality, considering measurements $z = 1,2$ are performed on party $C$; $b)$ given that party $C$ performed measurement $z = 3$, parties $AB$ evaluate the CHSH inequality corresponding to the result $c$.}
	\label{fig:diag1}
\end{figure}


\textit{Characterizing a specific measurement} - In the previous section, we have contented ourselves with trying to assess whether the measurement $C_3$ is entangling/entangled or not. However, the protocol that we defined can lead to a much finer statement. Indeed, if one is close to satisfying (\ref{cond2}), the measurement $C_3$ is close to an ideal Bell-state measurement. It should therefore be possible to bound the distance $t$ between the actual and the ideal measurement as a function the observed violations. The derivation of this bound in a full device-independent scenario, $t\leq f_{DI}(S)$, hits the same difficulties as those encountered in the simpler task of state estimation \cite{bardyn}. Here, we introduce additional assumptions and obtain a bound $t\leq f(S)$. Since obviously $f(S)\leq f_{DI}(S)$, we can conclude that a device-independent estimate of $t$ will be \textit{at least as bad as} $f(S)$.

We go back to the four-qubit scenario described after eq.~(\ref{cond2}) and we keep everything as there, except for the measurement $C_3$: this is no longer a perfect Bell state measurement, but is still assumed to be projective. One does not know \textit{a priori} which state to associate with each result $c_3$; however, once the measured data have been sorted out according to $c_3$, one can check all the four versions (\ref{versions}) of CHSH, and associate to each value of $c_3$ the version that leads to the maximal violation. This amounts at possibly relabeling the outcomes so that the eigenstate $\ket{e_c}$ is the closest to $\ket{\Phi_c}$ for each $c\in\{1,2,3,4\}$. We assume this to be the case from now on.

An operational measure of the distance between $C_3$ and an ideal Bell state measurement is the trace distance
\ba
t&=&\max_c\,\sqrt{1-| \langle e_c | \Phi _c \rangle |^2}\,.
\ea Note that this figure of merit represents the worst case scenario, since we are not specifying in which task the entangled measurement is going to be used after the check.

\begin{figure}[htbp!]
	\includegraphics[height = 50mm,width=65mm]{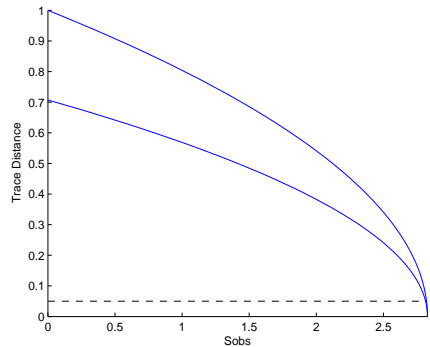}
	\caption{In the four-qubit scenario we consider the value of $S_{AB|c}$ as a device-assessment criterion, where the blue lines show the bounds on the trace distance of the measurment device. The violation $S_{obs} \geq 2.8214$ guarantees the trace distance of the measurement device to be at most 5\% from the ideal, as shown by the dashed line.}\label{figbounds}
\end{figure}

Now, because of the choice of the local measurements of Alice and Bob, the Bell operators corresponding to the four inequalities (\ref{versions}) read ${\cal B}_{AB|c}=2\sqrt{2}\left(\Phi_c-\Phi_{5-c}\right)$, where $\Phi_k=\ket{\Phi_k}\bra{\Phi_k}$. Therefore \ba
S_{AB|c} &=& 2\sqrt{2} ( | \langle e_c |\Phi _c \rangle |^2 -| \langle e_c |\Phi _{(5-c)} \rangle |^2 )\, ,
\ea
and the two bounds $0\leq| \langle e_c |\Phi _{(5-c)} \rangle |^2\leq 1-| \langle e_c |\Phi _c \rangle |^2$ lead finally to
\ba
\sqrt{\frac{1}{2} \big(1-\max_c \frac{S_{AB|c}}{2\sqrt{2}}\big)}&\leq t\leq & \sqrt{1-\min_{c} \frac{S_{AB|c}}{2\sqrt{2}}}\,.\label{eqbounds}
\ea In particular, the upper bound is the expression $f(S)$ we were looking for, and it allows us to assess how stringent are the requirements for device-independent assessment of a measurement. Indeed, recall that the trace distance is also the probability of distinguishing the real case from the ideal one \cite{trace dist}. Requesting that this probability is 5\% looks like a pretty loose requirement; but in order to confirm this assessment in a device independent way, one will have observe at least $\min_{c} S_{AB|c}\gtrsim 2.8214$ (Fig.~\ref{figbounds}). This number is within 0.5\% of the maximal value: no experiment has reached such a high violation and precision, even leaving aside that we are considering an entanglement swapping experiment.

\textit{Conclusion} - In the present letter we have presented a proposal for a device-independent test of an entangled measurement. Our proposal requires the use of three parties in an entanglement swapping scenario. There are several extensions and open problems which follow from our paper. In particular, our quantitative results rely on the fact that either Alice or Bob violate CHSH maximally with Charlie; it will be necessary to extend these results to less idealized situations.

One may also wonder if our three-partite scenario is the simplest one. While we do not have a definite answer, we conjecture that it would be impossible to achieve this task in a scenario involving only two parties.


\textit{Acknowledgements} - This work was supported by the National Research Foundation and the Ministry of Education, Singapore, and by the UK EPSRC. We thank the non-local club of CQT, as well as Nicolas Gisin, Miguel Navascu\'es, Stefano Pironio and Tamas V\'ertesi for enlightening comments.

\begin{appendix}

\textit{Appendix} - Given any two Hermitian operators $A_{0}, A_{1}$, with eigenvalues $\pm1$, acting on a Hilbert space $\HH$, there is a decomposition of $\HH$ into a direct sum of subspaces $\HH_{i}$ such that $\textrm{dim}\left(\HH_{i}\right) \leq 2 \, \, \forall \, \, i$ and both $A_{0}$ and $A_{1}$ act within each $\HH_{i}$, \textit{i. e.}, $\forall \, \ket{\psi} \in \HH_{i}, \, A_{0}\ket{\psi}, A_{1}\ket{\psi} \in \HH_{i}$ \cite{masanes}. Thus, the operators $A_{0}$ and $A_{1}$ can always be written as $A_{0} = \sum_{i} \Pi_{i}A_{0}\Pi_{i}$, $A_{1} = \sum_{i} \Pi_{i}A_{1}\Pi_{i}$, where $\Pi_{i}$ are projectors onto subspaces $\HH_{i}$. As a consequence, any Bell-CHSH operator $\beta$ acting on the Hilbert space of a 2-qudit system, $\HH = \CC^{d}\otimes\CC^{d}$, can be decomposed into a direct sum of Bell-CHSH operators $\beta_{i,j}$, each acting on a 2-qubit subspace \cite{foot} $\HH_{i,j}$, that is,
\ba
\beta & = & \oplus_{i,j} \beta_{i,j} = \nonumber \\ & = & \sum_{i,j} \left(\Pi_{i}\otimes\Pi_{j}\right)\beta\left(\Pi_{i}\otimes\Pi_{j}\right).
\ea

To evaluate $S_{Sep} = \textrm{max}_{\{\rho \in \mathcal{S}\}} \Tr{\left(\rho\beta\right)}$, where $\mathcal{S}$ is the set of separable states, we first note that, since the trace is linear  and $\mathcal{S}$ is a convex set, the maximum is attained over the subset of extremal points. Hence, it suffices to consider the set of pure product states $\mathcal{P}$. Now, considering this fact and the decomposition above stated, we have
\ba
S_{Sep} & = & \textrm{max}_{\{\ket{\phi} \in \mathcal{P}\}} \bra{\phi}\beta\ket{\phi} \nonumber = \\
& = & \textrm{max}_{\{\ket{\phi} \in \mathcal{P}\}} \bra{\phi}\oplus_{i,j}\beta_{i,j}\ket{\phi} = \nonumber \\ & = & \textrm{max}_{\{\ket{\phi_{i,j}} \in \mathcal{P}\}} \sum_{i,j} p_{i,j} \bra{\phi_{i,j}}\beta_{i,j}\ket{\phi_{i,j}},
\ea
where $\ket{\phi_{i,j}} = {\left( \Pi_{i}\otimes\Pi_{j}\right)\ket{\phi}}/{\sqrt{p_{i,j}}}$ and $p_{i,j} = \bra{\phi}\left( \Pi_{i}\otimes\Pi_{j}\right)\ket{\phi}$. By convexity, the above maximum is upper bounded by the largest mean value among the 2-qubit Bell operators $\beta_{i,j}$ attained by 2-qubit pure product states:
\ba
S_{Sep} & = & \textrm{max}_{\{\ket{\phi_{i,j}} \in \mathcal{P}\}} \sum_{i,j} p_{i,j} \bra{\phi_{i,j}}\beta_{i,j}\ket{\phi_{i,j}} \leq \nonumber \\
& \leq & \sum_{i,j} p_{i,j} \, \textrm{max}_{\{\ket{\phi_{i,j}} \in \mathcal{P}\}} \bra{\phi_{i,j}}\beta_{i,j}\ket{\phi_{i,j}} \leq \nonumber \\
& \leq & \textrm{max}_{\{\ket{\phi} \in \mathcal{P}, (i,j)\}} \bra{\phi}\beta_{i,j}\ket{\phi}.
\ea
According to \cite{scarani}, any 2-qubit Bell-CHSH operator has the following spectral decomposition, up to local unitaries,
\ba
\beta & = & \sum_{i = 1}^{4} \alpha_{i} \ket{\Phi_{i}}\bra{\Phi_{i}},
\ea
where the eigenvectors $\ket{\Phi_{i}}$ are Bell states and the eigenvalues are functions of the local observables, constrained to $\alpha_{1} = -\alpha_{3}$, $\alpha_{2} = -\alpha_{4}$, $\alpha_{1}^2 + \alpha_{2}^2 = 8$. Let $\alpha_{i,j}$ be the largest eigenvalue of $\beta_{i,j}$. We have
\ba
S_{Sep} & = & \textrm{max}_{\{\ket{\phi} \in \mathcal{P}, (i,j)\}} \bra{\phi}\beta_{i,j}\ket{\phi} = \nonumber \\
& = & \textrm{max}_{\{\ket{\phi} \in \mathcal{P}, (i,j)\}} \, \alpha_{i,j} \left|\braket{\phi}{\Phi_{1}}\right|^2 + \nonumber \\ & + & \sqrt{8 - \alpha_{i,j}^2} \left|\braket{\phi}{\Phi_{2}}\right|^2 -  \alpha_{i,j} \left|\braket{\phi}{\Phi_{3}}\right|^2 - \nonumber \\ & - & \sqrt{8 - \alpha_{i,j}^2} \left|\braket{\phi}{\Phi_{4}}\right|^2 .
\ea
Without loss of generality, we do not worry about the local unitaries in the spectral decomposition of $\beta$ since they can be absorbed into the states $\ket{\phi}$. The best projection of a pure product state into a Bell state has probability $1/2$; this allows us to perform the optimization over the states,
\ba
S_{Sep} & = & \textrm{max}_{\{(i,j)\}} \frac{\alpha_{i,j}+\sqrt{8-\alpha_{i,j}^2}}{2}.
\ea

It is important to note, at this point, that, for all $(i,j)$, $\alpha_{i,j} \geq 2$. This is so because the largest eigenvalue $\alpha$ of $\beta$ is given by the positive square root of the largest eigenvalue of $\beta^2$, which is lower bounded by $2$ \cite{landau}. We observe that the above function decreases as $\alpha$ increases. This way, the maximum is attained for the subspace $(i,j)$ such that $\alpha_{i,j}$ is minimum. Then, defining $\lambda$ as the smallest eigenvalue of $\beta$ such that $\lambda \geq 2$, we have
\ba \label{lemma}
S_{Sep} & = & \frac{\lambda+\sqrt{8-\lambda^2}}{2}.
\ea

By means of a simpler treatment, this result is in complete accordance - and generalizes to higher dimensions - the results of \cite{seevinck, roy}. We now state and prove our main result.

\textit{Theorem}:
If $S_{AC} = S_{BC} = 2\sqrt{2}$, and if $C_3$ is a separable measurement, then $S_{AB|c_3}\leq \sqrt{2}$.

\textit{Proof of theorem} - As previously stated, if $S_{AC} = S_{BC} = 2\sqrt{2}$, then, for all subspaces $(i,j)_{AC}$ and $(k,l)_{BC}$ where the initial states shared by parties $AC$ and $BC$ have support on, the states are, up to local isometries, maximally entangled states and are completelly uncorrelated from any other system. This way, any state steered by measurement $C_{3}$ - assuming it is separable - to parties $AB$ will be product, and will have support at most in the same subspaces of $\HH_{A}$ and $\HH_{B}$ where the initial states have support. Moreover, implicit in $S_{AC} = S_{BC} = 2\sqrt{2}$ is the statement that in every subspaces $(i,j)_{AC}$ and $(k,l)_{BC}$ where the initial states have support the Bell-CHSH operators $\beta_{i,j}$ and $\beta_{k,l}$ have maximal eigenvalues $\alpha_{i,j} = \alpha_{k,l} = 2\sqrt{2}$. This immediatly implies that, for the same subspaces, the Bell-CHSH operators in parties AB, $\beta_{i,k}$, will also have maximal eigenvalues $\alpha_{i,k} = 2\sqrt{2}$; this follows from  \cite{landau}.

Thus, we conclude that, for all subspaces $(i,k)$ where the final steered (separable) state in parties $AB$ has support, the 2-qubit Bell-CHSH operators have maximum eigenvalue $2\sqrt{2}$. Hence, by (\ref{lemma}), we finally have
\ba
S_{AB|c_3}\leq \sqrt{2}.
\ea

\end{appendix}


\begin{thebibliography}{99}
\bibitem{foot1} {To see this, just notice that one possible way of decreasing the violation between Alice and Charlie consists in keeping the state $\Phi_{AC}\otimes\rho_{A'BC'}$ but change the measurement so that it starts involving the degrees of freedom $A'$. There is no way of excluding entanglement between $A'$ and Bob's system.}
\bibitem{foot} Note that this is true only if the dimension of the Hilbert space is even; if it is odd, despite the 2-qubit Bell operators there will be, on the decomposition, one more term. This extra term plays no role in the following calculations. Hence, without any loss, we assume the dimension of the Hilbert space is even.
\bibitem{note} Note that, by convexity, the maximum value of $S_{AB}$ will be attained by pure states, which are extremal.
\end{thebibliography}
\end{document}